%% file: paper_continuum.tex
\documentclass[useAMS,usenatbib]{mnras}
\usepackage{amssymb,amsmath,epsfig,times,natbib,longtable,color,graphicx,dcolumn,txfonts,textcomp}
\input{journal_defs}

\newcommand{\nustar}{\textit{NuSTAR}}

\newcommand{\swift}{{\it Swift}}
\newcommand{\xmm}{{\it XMM-Newton}}

\newcommand{\hitomi}{{\it Hitomi}}

\makeatletter
 \def\hlinewd#1{%
   \noalign{\ifnum0=`}\fi\hrule \@height #1 \futurelet
    \reserved@a\@xhline}
\makeatother

\title[Continuum fitting with high energy detectors]{XRB continuum fitting with sensitive high energy X-ray detectors}

\author[M. L. Parker et al.]{M. L. Parker$^{1,2}$,\thanks{Email: mparker@sciops.esa.int}
  D. J. K. Buisson,$^2$
  J. A. Tomsick,$^{3}$
  A. C. Fabian,$^{2}$
  K. K. Madsen,$^{4}$\newauthor
  D. J. Walton$^{2}$
  and F. F{\"u}rst$^{1}$\\
  $^{1}$European Space Agency (ESA), European Space Astronomy Centre (ESAC), E-28691 Villanueva de la Ca\~{n}ada, Madrid, Spain\\
  $^{2}$Institute of Astronomy, Madingley Road, Cambridge, CB3 0HA, UK\\               
  $^{3}$Space Sciences Laboratory, University of California, Berkeley, 7 Gauss Way, Berkeley, CA 94720-7450, USA\\          
  $^{4}$Cahill Center for Astronomy and Astrophysics, California Institute of Technology, 1200 E. California Blvd, Pasadena, CA 91125, USA\\
}
\date{}

\begin{document}

\maketitle

\begin{abstract}
The launch of the \emph{Nuclear Spectroscopic Telescope Array} (\nustar ) heralded a new era of sensitive high energy X-ray spectroscopy for X-ray binaries (XRBs). In this paper we show how multiple physical parameters can be measured from the accretion disk spectrum when the high-energy side of the disk spectrum can be measured precisely using \nustar . This immediately makes two exciting developments possible. If the mass and distance of the source are known, the continuum fitting method can be used to calculate the spin and inner disk inclination independently of the iron line fitting method. If the mass and distance are unknown, the two methods can be combined to constrain these values to a narrow region of parameter space. In this paper we perform extensive simulations to establish the reliability of these techniques. We find that with high quality spectra, spin and inclination can indeed be simultaneously measured using the disk spectrum. These measurements are much more precise at higher spin values, where the relativistic effects are stronger. The inclusion of a soft X-ray snapshot observation alongside the \nustar\ data significantly improves the reliability, particularly for lower temperature disks, as it gives a greatly improved measurement of the disk peak. High signal to noise data are not necessary for this, as measuring the peak temperature is relatively easy. We discuss the impact of systematic effects on this technique, and the implications of our results such as robust measurements of accretion disk warps and XRB mass surveys.
\end{abstract}


\section{Introduction}

There are two main methods for establishing the spin of a black hole X-ray binary. The first, the iron line or reflection method, relies on measuring the relativistic distortion of the narrow iron K$\alpha$ line induced by the rotation of the inner disk and by the gravitational redshift of the black hole \citep{Fabian89,Reynolds14}. The second method, continuum fitting, relies on accurately establishing the luminosity of the disk spectrum, which is a strong function of the dimensionless spin parameter $a$ \citep{Zhang97,McClintock14}. This method is generally only applicable when the mass, $M$, and distance, $D$, of the black hole concerned are known, as otherwise the luminosity and baseline temperature of the disk are unknown. The iron line method thus relies on precise measurement of the spectral shape and uses no information about the absolute flux in the line, and the continuum fitting method largely relies on the flux and uses relatively little spectral information, due to the far smoother shape of the disk spectrum.

Both of these methods have been used extensively to measure the spin of X-ray binaries. \citet{Middleton16} updates the table of \citet{Miller_and_Miller15} and lists 5 XRBs with continuum fitting spins, 8 with reflection spins, and 6 with both. Generally, the two techniques are applied to different spectra: continuum fitting requires a strong disk component, which can obscure or confuse the iron line measurements used in reflection spectroscopy. Because of this, the two methods are rarely used simultaneously, which limits how comparable their results are, as the accretion geometry may change between different observations. Despite this, the agreement between the two methods is generally good - only two sources have significant differences: 4U~1543$-$475 ($a_\mathrm{ref}=0.3\pm0.1$, $a_\mathrm{cont}=0.8\pm0.1$) and GRO~J1655$-$40 ($a_\mathrm{ref}=0.98\pm0.01$, $a_\mathrm{cont}=0.7\pm0.1$). While it may be the case that this is just attributable to statistical or systematic errors, or truncation of the disk in some states, this may also be a signature of incorrect assumptions in the modeling, which can be used to probe accretion physics.

Because of the broad spectral shape of the disk it is much harder to measure the relativistic distortion of the spectrum, and instead spin measurements using continuum fitting rely on measuring the temperature and luminosity of the disk \citep[see reviews by][]{Remillard06,McClintock14}. Both of these parameters are strongly affected by the spin, which controls the radius of the innermost stable circular orbit (ISCO). The bulk of the high energy emission from the disk comes from the innermost radii, where the disk is hottest and the radiation can be strongly relativistically boosted. However, this also depends on the inclination of the disk with respect to the observer, as this controls the level of boosting and Doppler shifting of the emitted radiation. Again, this is most prominent at small radii, where the orbital velocities are largest. This degeneracy is broken by assuming that the inclination of the inner disk is the same as that of the binary system, which can be established along with $M$ and $D$ by using dynamical measurements \citep[see e.g.][and references therein]{Ozel10}.
This technique has successfully been applied to a large number of spectra from several different sources \citep[e.g.][]{Shafee06, McClintock06,Liu08,Steiner09,Steiner10,Steiner11,Gou09,Gou11}

However, the requirement for a known mass, distance and inclination limits the application of the continuum fitting method to the handful of sources where these parameters are known, and relies on the assumption that the disk and binary inclinations are the same.
In principle, however, this limitation can be partly overcome with sufficiently high quality data. While there is a strong degeneracy between $a$ and $i$, it is not absolute. This means that with enough signal $a$ and $i$ can be measured simultaneously. This necessitates the use of precision spectroscopy with a highly sensitive instrument, which covers the high energy side of the disk spectrum.

\nustar\ \citep{Harrison13} is the first and currently only focusing hard X-ray telescope, although it was briefly joined by the hard X-ray telescope (HXT) aboard \hitomi\ \citep{Kokubun12}. The \nustar\ detectors are uniquely suited to observations of X-ray binaries due to the absence of the pileup effects that plague CCD detectors at high count rates \citep[when multiple photons arrive within a single read-out time and are interpreted as one event, see e.g.][]{Miller10}. By reading each event as it occurs, rather than reading all events with a fixed frequency, this problem is eliminated. This means that highly sensitive measurements of the broad iron line have been made possible by \nustar\ \citep{Miller13, Miller13Serpens, Tomsick14, King14, Fuerst15, Parker15_cygx1, Parker16_gx339, Walton16_cygx1, Walton17_v404}. 

This paper is a follow-up to \citet{Parker16_gx339}, hereafter P16, where we demonstrated the potential of \nustar\ to constrain multiple disk parameters simultaneously. In this work, we will examine the suitability of continuum fitting to constrain both spin and (inner disk) inclination at the same time, using simulations. In section~\ref{sec_method} we describe the basic principle of fitting the disk spectrum for multiple parameters. In section~\ref{sec_simulations} we simulate a large number of spectra to test the reliability of the method. In section~\ref{sec_errors} we explore some potential sources of systematic error in this technique. Our discussion and conclusions are in sections~\ref{section_discussion} and~\ref{section_conclusions}.

\section{Principle of the method}
\label{sec_method}

\begin{figure}
\centering
\includegraphics[width=\linewidth]{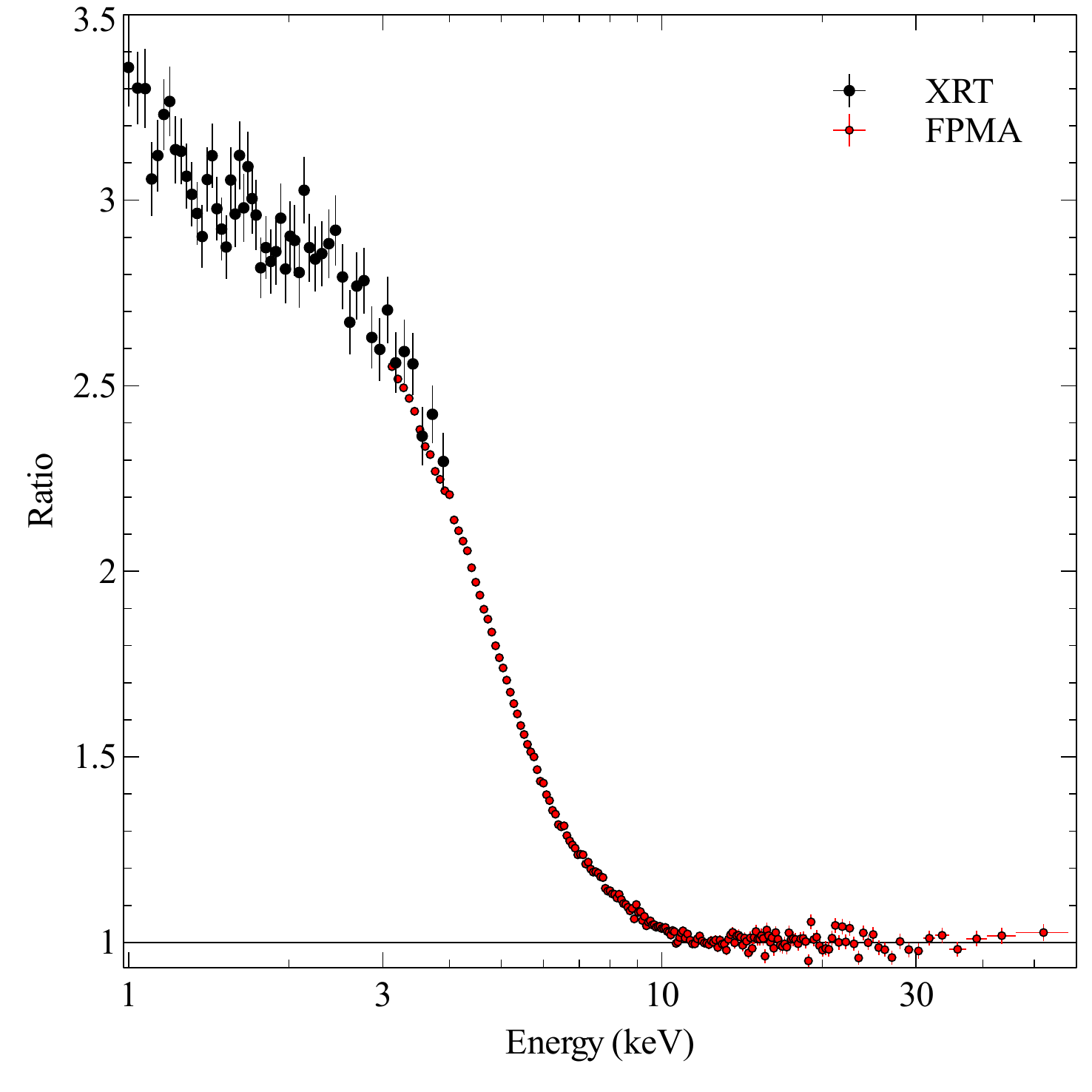}
\caption{The data/model ratio from the GX~339-4 very high state \nustar\ and \swift\ observation P16, with the \emph{kerrbb} disk model removed. These residuals show the disk contributes a significant amount of flux up to 10~keV. Note that we only show the spectrum from one of the two \nustar\ modules, so twice as much data is available. The peak of the disk spectrum is around 1~keV, but the disk fraction continues to rise below this point due to the drop in photons in the Compton-scattered tail.}
\label{fig_gx339_ratio}
\end{figure}

In Fig.~\ref{fig_gx339_ratio} we show the data/model residuals of the very high state spectrum of GX~339-4 to the best-fit model presented in \citet{Parker16_gx339}, with the relativistic disk spectrum removed. It is clear from this figure that the high energy side of the disk is contributing a significant amount of flux well into the \nustar\ band (up to $\sim$10~keV), meaning that small distortions in the spectral shape can potentially be measured, given the high quality of the \nustar\ data (only one of the two FPM spectra is shown, and below $\sim$15~keV the error bars are smaller than the points). 

\begin{figure*}
\centering
\includegraphics[width=\linewidth]{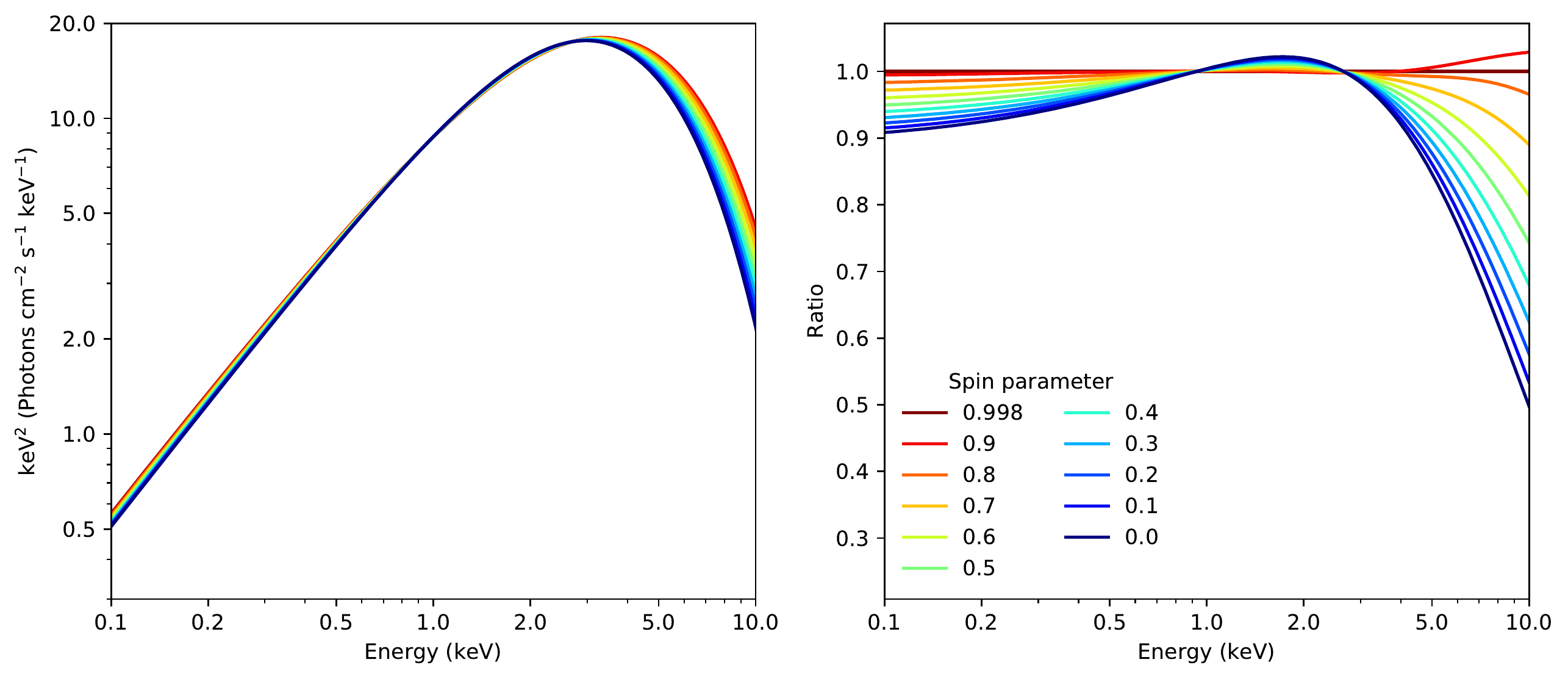}
\caption{Left: \textsc{kerrbb} models with fixed $a$ fit to a simulated spectrum with $a=0.998$. Below $\sim3$~keV the spectra are almost completely degenerate, but at high energies large differences start to appear. Right: as left, but as a ratio to the original $a=0.998$ model. At 10~keV, 50\% of the model flux is lost by fitting with $a=0$. This can easily be measured, provided that the disk spectrum can still be constrained at high enough energies.}
\label{fig_kerrbb_li}
\end{figure*}

To explore how this high sensitivity can be exploited, we consider a relativistic disk spectrum, using the \textsc{kerrbb} model \citep{Li05_kerrbb}. We simulate an idealised 0.5--10~keV spectrum, with no background, no noise, a 100~ks exposure time and the \xmm\ instrument response, using \textsc{kerrbb} with parameters $M=10M_\odot$, $D=10$~kpc, $a=0.998$, $i=40$\textdegree and $\dot{M}=0.2\dot{M}_\mathrm{Edd}$, where $\dot{M}_\mathrm{Edd}$ is the Eddington accretion rate. We then re-fit this data, stepping $a$ from 0.998 to 0, with $M$ and $D$ fixed and the other parameters free to vary. In this case, as in all others, we keep the normalization of \textsc{kerrbb} fixed to 1. The results of this are shown in Fig.~\ref{fig_kerrbb_li}. It is obvious that at low energies the spectra are extremely similar. However, at higher energies there are very large differences between the models (50\% less flux at 10~keV in the $a=0$ case). These differences are largely due to the increased high energy flux for rapidly rotating black holes, caused by Doppler shifting and relativistic beaming of the photons coming from the approaching side of the disk. At higher spin values, the velocities at the ISCO are larger, meaning that the beaming is stronger. This can be mimicked to an extent by increasing the viewing inclination (with respect to the disk axis, so that the disk is more edge-on to the observer) giving a higher line-of sight velocity. However, this does not affect the absolute velocity of the disk, and so cannot fully reproduce a higher spin spectrum.

In relative terms, the differences in spectral shape measured using this method are small, due to the rapid drop in disk flux at high energies, and can only be reliably measured because of the very high sensitivity of \nustar\ or similar instruments. To demonstrate this, we consider a more realistic simulated \nustar\ spectrum from our sample described in \S~\ref{sec_a_and_i}, with a power-law tail, where the spin is well constrained. We take the spectrum for $M=10M_\odot$, $D=3$~kpc, $a=0.998$, and $i=40$\textdegree, and manually fit it. We then re-fit with the spin fixed at different values to determine the effect on the spectrum. The results of this are given in Table~\ref{table_wrongspinfits}. While this is a simplified spectrum, the result is clear: while there is a degeneracy between spin and inclination, a high spin relativistic disk spectrum cannot be approximated by a low spin disk spectrum to arbitrary precision, and the level of spectral precision required to break this degeneracy is well within the reach of current instrumentation. There is therefore no reason, in principle, why spin and inclination cannot be simultaneously measured.

\begin{table}
\centering
\caption{Estimated parameters and fit statistics for fitting a spectrum with the spin fixed at different values.}
\label{table_wrongspinfits}
\begin{tabular}{l c c c r}
\hline
\hline
Parameter&$a$ & $i$ & $\dot{M}$ & $\chi^2/\mathrm{dof}$\\
&&(degrees) & (10$^{18}$~g~s$^{-1}$)\\
\hline
Simulated&0.998 & 40 & 2.74\\
\hline
Best-fit &$>0.98$ & $43_{-3}^{+2}$ & $3.2_{-0.4}^{+0.2}$ & 383/371\\
\hline
Fixed 	& $0.9$ & $54.70\pm0.03$ 			& $4.770\pm0.001$ 			& 418/372\\
		& $0.8$ & $63.52\pm0.03$			& $6.711\pm0.002$			& 465/372\\
		& $0.7$ & $70.36\pm0.03$			& $9.400\pm0.005$			& 504/372\\
		& $0.6$	& $75.70\pm0.03$			& $13.21\pm0.01$			& 517/372\\
		& $0.5$ & $79.74\pm0.03$			& $18.53\pm0.02$			& 519/372\\
		& $0.4$ & $82.71\pm0.02$			& $25.71\pm0.04$			& 526/372\\
		& $0.3$ & $84.88\pm0.02$			& $35.06\pm0.05$			& 567/372\\
\hline
\hline
\end{tabular}
\end{table}

The most obvious application of this is to free the inclination parameter, allowing it to differ from that of the binary system. Of the three fixed parameters, this is the only one that could theoretically differ from that obtained from dynamic measurements. When the black hole spin axis and the binary orbit are not aligned, the inner disk is likely to warp \citep{Bardeen75}, aligning itself with the black hole spin, while the outer disk remains aligned with the binary system. There have been hints of differences in inclination between the dynamic and reflection measurements in some sources. For example, inclinations measured using reflection spectroscopy of Cyg~X-1 with \nustar\ \citep{Tomsick14,Parker15_cygx1,Walton16_cygx1} have generally been ~10--15\textdegree\ higher than that inferred of the binary system \citep{Orosz11}.  Using this technique we can now obtain an independent measurement of the inclination using continuum fitting, giving three independent measurements of the inclination (binary, reflection, and continuum). If those obtained from the X-ray spectra (which only measure the inner disk) systematically differ from the binary inclination (which determines the outer disk inclination), we can be much more confident of a warp in the disk.

%
A second, less obvious, application is to combine the continuum fitting and reflection methods to swap which parameters are fixed and which are measured. Instead of using $M$ and $D$ as the input for the continuum fitting model, the $a$ and $i$ values obtained with reflection can be used, and $M$ and $D$ can be constrained. This method has two main sources of error: the intrinsic uncertainty in measurements of $M$ and $D$, due to the degeneracy between the two and the accretion rate, and the uncertainty inherited from the $a$ and $i$ values from reflection. How reliably this method can be used depends on the relative strengths of these effects - ideally, the intrinsic uncertainty will be small, so the net error is dominated by the uncertainty in $a$ and $i$ from reflection, which can be measured relatively accurately We explore this in section~\ref{sec_MD}.

\section{Simulations}
\label{sec_simulations}

\subsection{Spin and inclination}
\label{sec_a_and_i}

\begin{figure*}
\centering
\includegraphics[width=14cm]{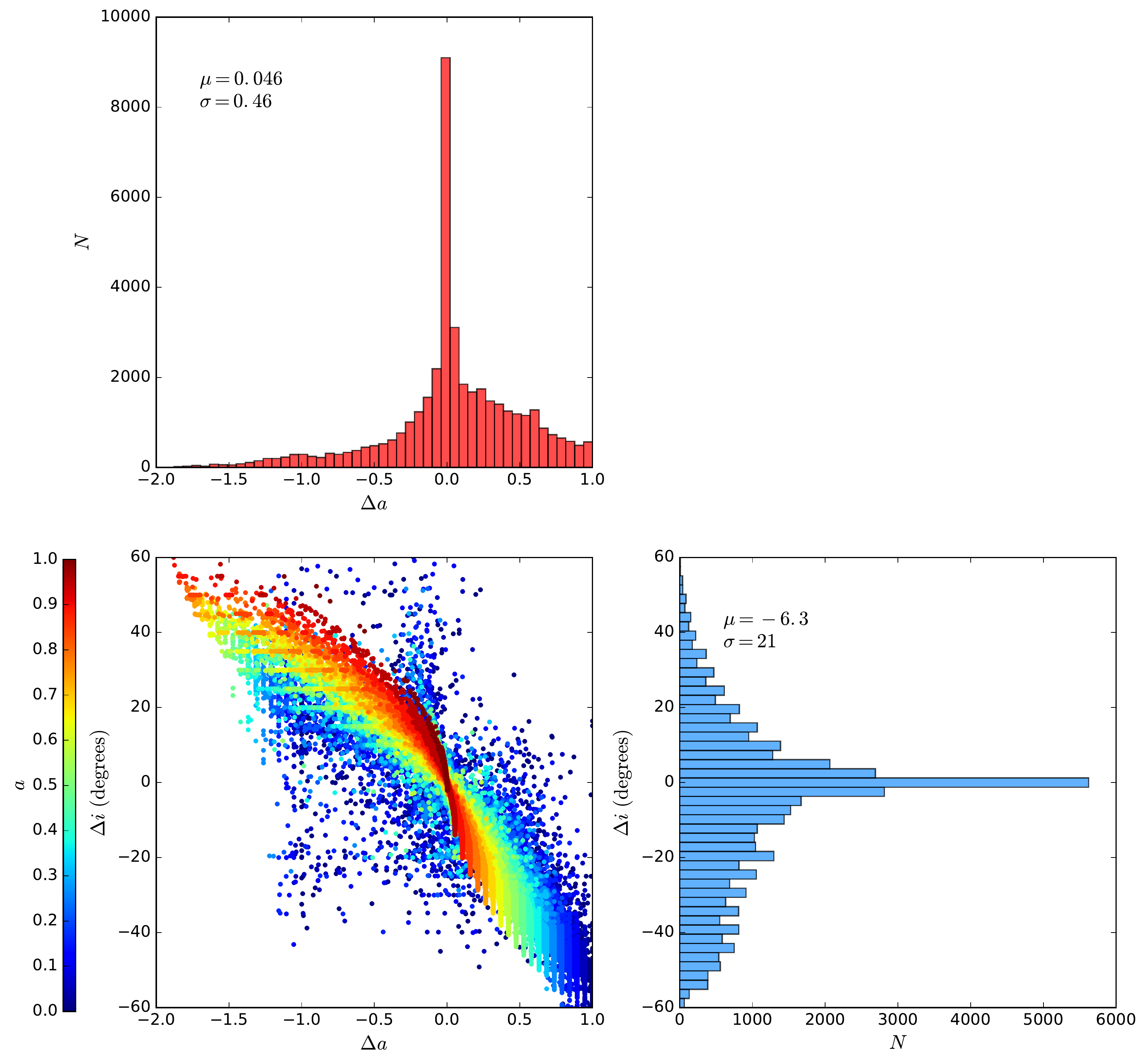}
\caption{The spin-inclination degeneracy for all 41600 spectra in the sample. The lower left plot shows the difference between the measured and simulated values of $a$ and $i$, colour-coded by the spin, and the two histograms show the distributions in $\Delta a$ and $\Delta i$ after marginalising over the other parameter. $\mu$ and $sigma$ are the mean and standard deviation, for each parameter. The series of peaks in the inclination histogram are caused by fits hitting the edge of the allowed range of inclinations.}
\label{fig_sims_spinincdegeneracy}
\end{figure*}

To test the accuracy of fitting spin and inclination simultaneously we simulate a grid of spectra with different $a$, $i$, $M$ and $D$ values and then fit the model back to the data. For this test, we use a simple model of a relativistic disk (\textsc{polykerrbb}, a version of \textsc{kerrbb} modified to allow high order interpolation, see Appendix~\ref{sec_polykbb}) and Comptonized tail (\textsc{simpl}), modified by absorption (\textsc{tbabs}) with a column density of $5\times10^{21}$~cm$^{-1}$. We simulate 41600 spectra, with 16 values of $M$ from 5 to 20~$M_\odot$, 10 values of $D$ from 1 to 10~kpc, 20 values of $a$ from 0 to 0.998, and 13 values of $i$ from 20\textdegree\ to 80\textdegree. All grid parameters are linearly spaced. The photon index of \textsc{simpl} is set to 2, and the fraction of photons scattered into the power-law tail is 0.1. At each step on the grid, the accretion rate is adjusted to $0.2\times\dot{M}_\textrm{Edd}$.
We simulate a long 100~ks \nustar\ exposure\footnote{Note that for our purposes, exposure time is entirely degenerate with distance, and thus the effect of shorter exposures will be identical to the effect of larger distances.} for each spectrum, using the official background and response files for simulations\footnote{\url{https://www.nustar.caltech.edu/page/response_files}}. We then fit these spectra with the same model, fixing the mass and distance at each step but allowing all other parameters (including those of \textsc{simpl}) to vary freely. We then record the final best fit spin and inclination estimates, and their errors. Additionally, we record the 3-10~keV flux of the disk component so that we can control for the effect of flux changes caused by the various parameters.

Due to the large number of spectra we use, the procedure for simulating and fitting had to be automated. We use the \textsc{pyxspec} interface for \textsc{xspec}, which allows it to be called from \textsc{python}. We use two scripts, one which simulates the spectra based on the specified parameters, and one which fits the resultant fake data. We manually verified a small number of randomly selected fits to ensure that the program was functioning as expected. Each simulation or fit was run individually, distributed over the X-ray cluster in Cambridge using \textsc{HTCondor}.
For the sake of speed, we simulate only one FPM spectrum, but double the exposure time to compensate. We rebin each spectrum to have a signal-to-noise ratio of 50 in every bin, and to oversample the data by a factor of 3. We ensure that for both simulating and fitting the energy grid is extended sufficiently far beyond the boundaries of the \nustar\ response that all models function as intended.

When fitting, we first set the parameters to those of the simulation, then run the standard \textsc{xspec} fit command. After the fit completes, we calculate the errors on $a$ and $i$ (frequently finding a new best fit in the process) using the standard \textsc{xspec} algorithm.

Fig.~\ref{fig_sims_spinincdegeneracy} shows the difference between the simulated and measured values of $a$ and $i$ for all 41600 spectra ($\Delta a$ and $\Delta i$, respectively). The distributions of $a$ and $i$ are both strongly peaked at the correct value ($\Delta a=\Delta i =0$). For a given spin value, indicated by a single colour, the points are tightly confined to a curve showing the degeneracy between $a$ and $i$, with some slight differences caused by different $M$ value. The lack of scatter about the curves indicates how strong a constraint can be obtained by \nustar\ alone on a single parameter - if the inclination is fixed, then the spin can be measured very precisely.
It is obvious from this figure that the constraints are stronger at high spin, with the points spread around the degeneracy curve at low spin and tightly clustered at maximal spin. This is caused by three separate effects: the disk extends further into the \nustar\ bandpass at high spin; the changes in spectral shape are much more pronounced; and the signal to noise is higher due to high spin spectra being brighter.

\begin{figure*}
\centering
\includegraphics[width=0.48\linewidth]{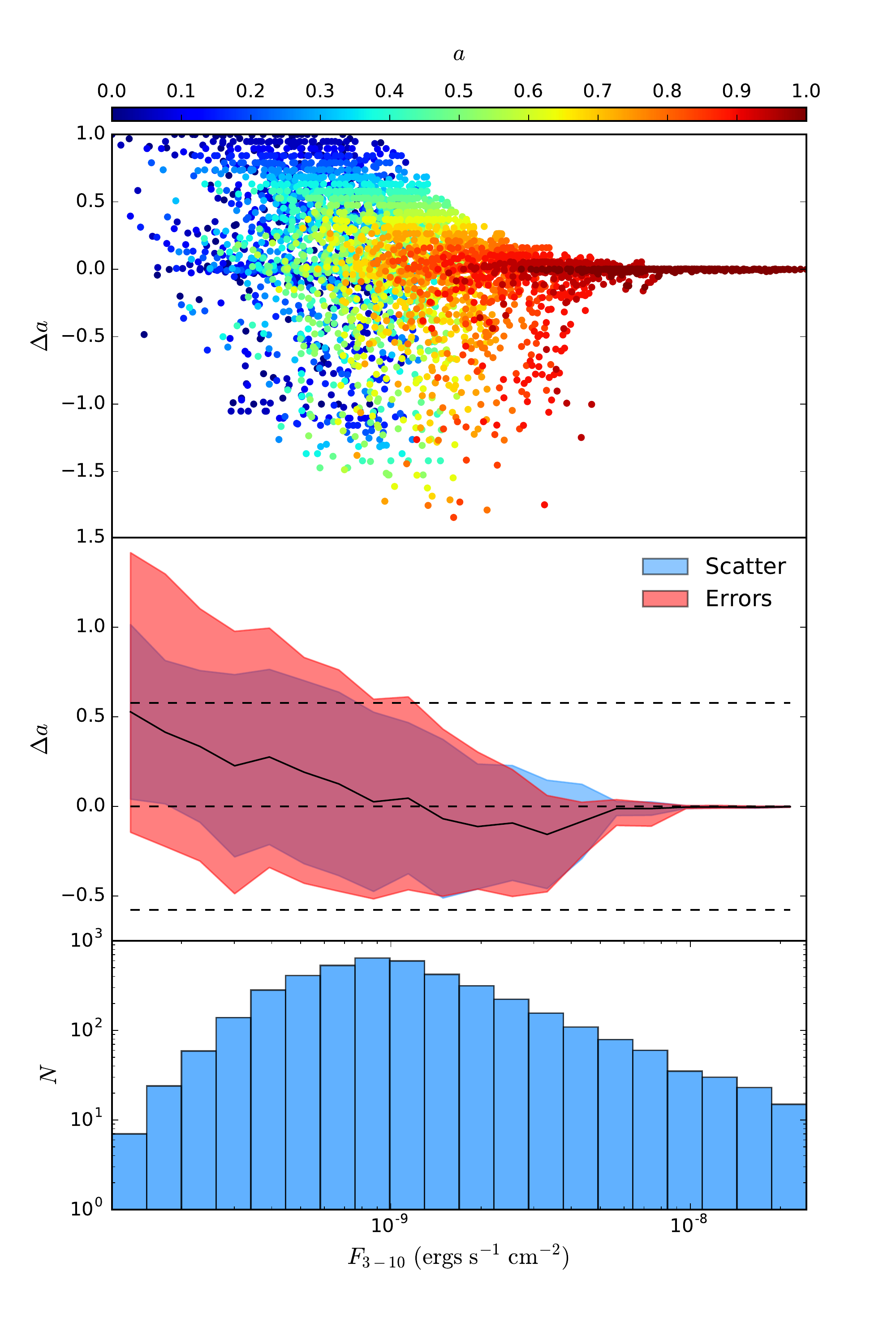}
\includegraphics[width=0.48\linewidth]{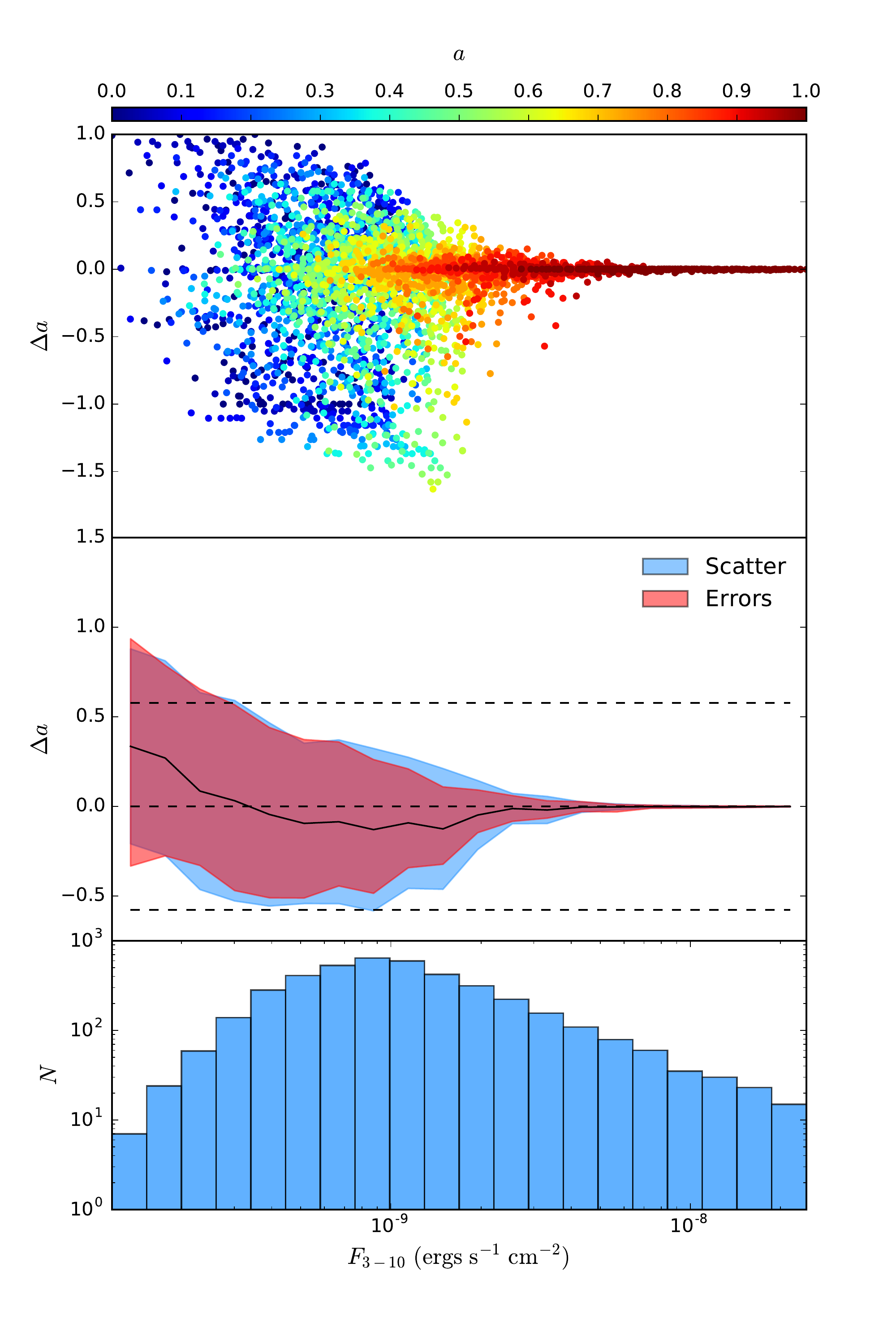}
\caption{\emph{Left:} The accuracy of the spin measurements as a function of flux. Top panel: all 4160 points for $D=5$~kpc, colour-coded by spin. Middle: the scatter in those points (blue) and the mean error for each point (red). Dashed horizontal lines indicate the standard deviation expected from a uniform distribution between $\pm1$ (i.e. if the spin was unconstrained and selected at random). Bottom: histogram of the number of points in each flux bin. \emph{Right:} As left, but for joint \nustar\ XRT fits.}
\label{fig_flux_dispersion}
\end{figure*}

In the left panel of Fig.~\ref{fig_flux_dispersion} we show the difference between the simulated and measured $a$ values, plotted as a function of 3--10~keV flux, for all spectra at $D=5$~kpc. There is a clear trend with flux: the higher the flux, the better the constraint obtained\footnote{Note that, in the case where the background is low, the source flux is interchangeable with exposure time or the effective area of the instrument used.}. Below around $10^{-9}$~ergs~s$^{-1}$~cm$^{-2}$, the scatter is around the magnitude expected for randomly chosen values, and there is a systematic bias towards higher spin. However, this increased uncertainty at low fluxes is reflected in the increased size of the errors (shown in the middle panel of Fig.~\ref{fig_flux_dispersion}). Based on this, we conclude that the measured errors on $a$ (and hence $i$, as the two are strongly correlated) are likely to be accurate, so the true spin value is likely to be within the measured error interval.

\begin{figure*}
\centering
\includegraphics[width=0.48\linewidth]{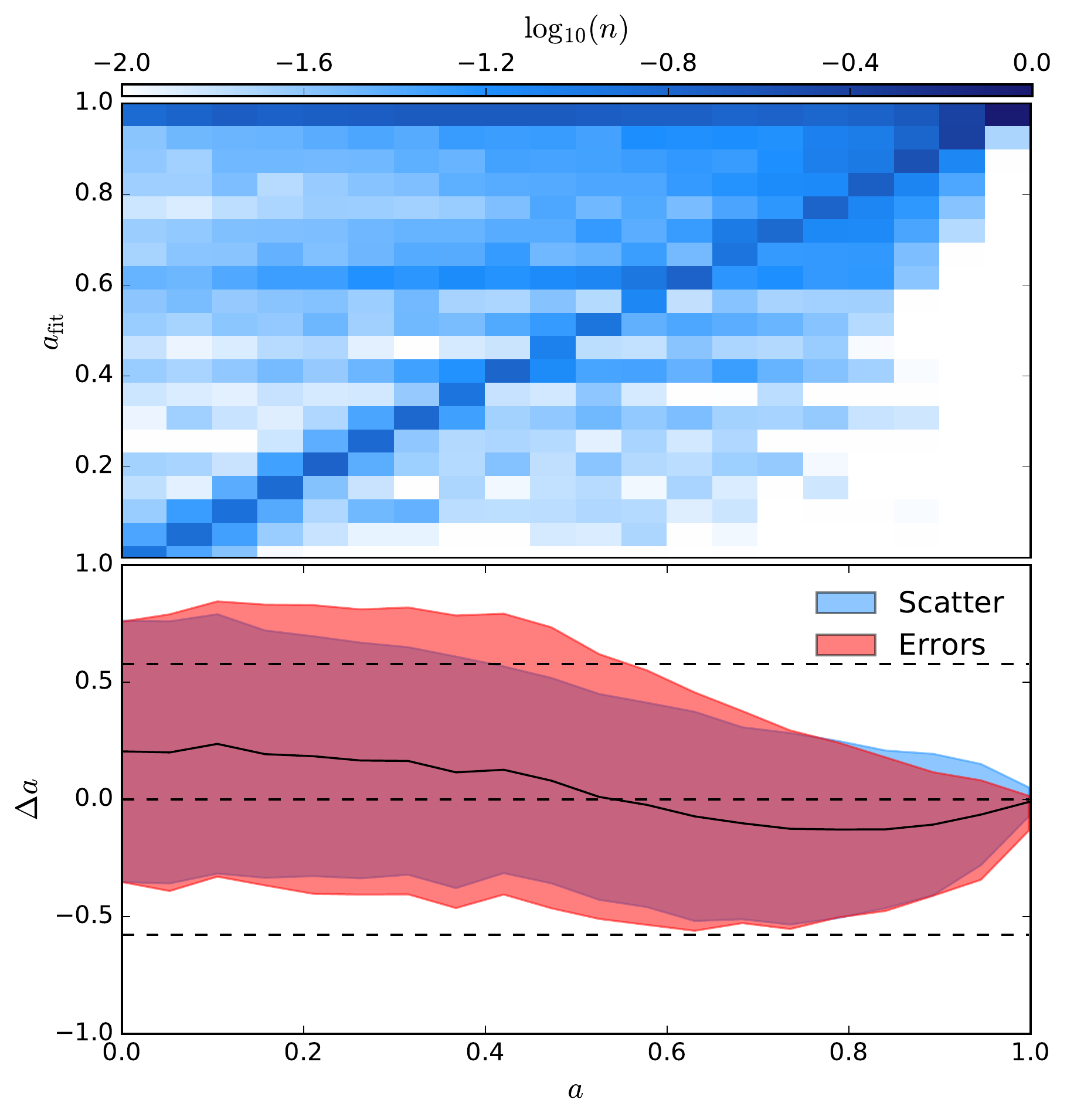}
\includegraphics[width=0.48\linewidth]{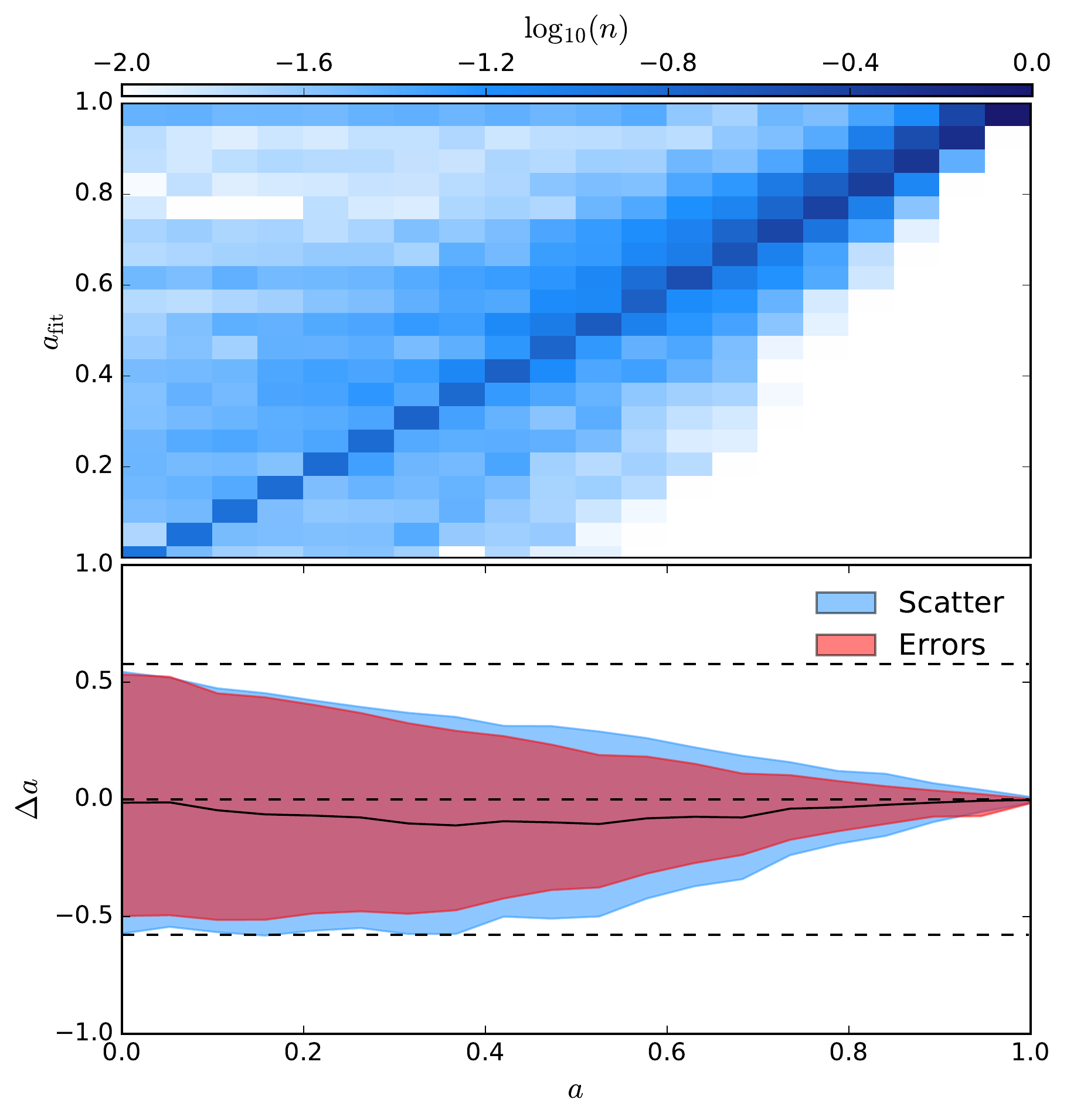}
\caption{\emph{Left:} The accuracy of the spin measurements as a function of spin. Top panel: 2D histogram, showing the spin measurements for all 41600 spectra as a function of spin. There is a strong peak at the correct value for all spins, but a significant number hit the upper limit for all values of spin. Bottom panel: The standard deviation of the measured spins as a function of simulated spin, and the mean $1\sigma$ error (calculated with the \textsc{xspec} error command) as a function of spin. The two are a good match, implying that the \textsc{xspec} errors are reliable. \emph{Right:} As left, but for the case where a \swift\ XRT snapshot is used in addition to the \nustar\ spectrum. The correlation is tighter, less biased, and has smaller scatter. Most notably, far fewer measurements hit the upper limit.}
\label{fig_spin_dispersion}
\end{figure*}

We also investigate the uncertainty in the measured spin as a function of spin (Fig.~\ref{fig_spin_dispersion}, left). The top panel is a 2D histogram, showing the distribution of measured spins as a function of spin. For all values of spin, there is a peak at the correct value, but there are also many points clustered at the parameter limits. Generally, the higher the spin, the more accurate the measurement of spin. Similarly to Fig.~\ref{fig_flux_dispersion}, the measurement errors accurately reflect the scatter in the values, meaning that measured values are likely to be reliable. 

\begin{figure*}
\includegraphics[width=14cm]{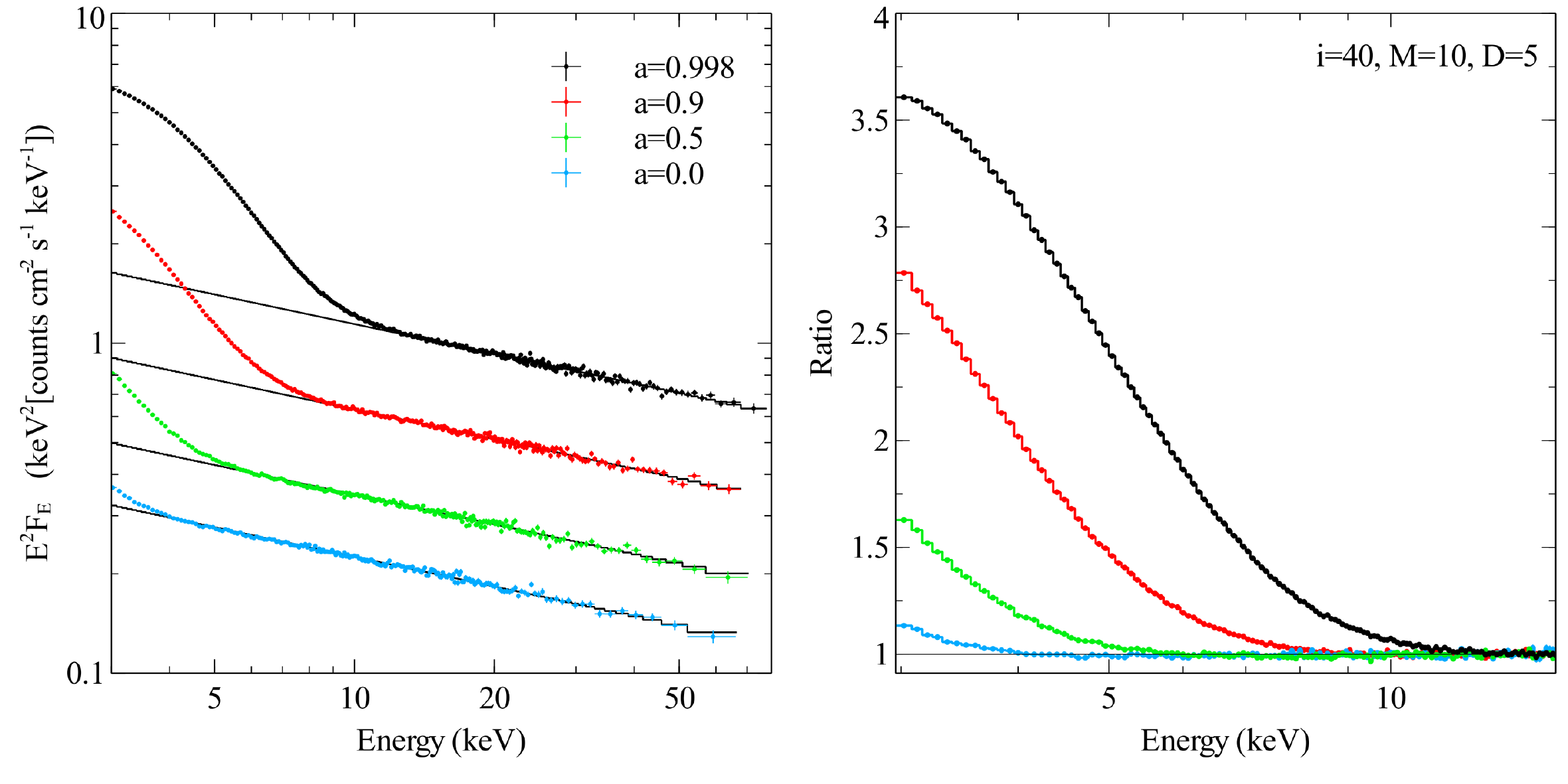}
\caption{Left: Unfolded disk plus Comptonization spectra for a range of spin values at the same accretion rate, taken from our simulated spectra. Data are fit with a power-law in the 20--79~keV band, which is then extrapolated to lower energies. Right: Ratio of the data to the extrapolated power-law below 20~keV.}
\label{fig_spin_ratios}
\end{figure*}

A secondary effect of increased spin, which is particularly important for \nustar\ data, is that it shifts the peak of the disk spectrum to higher energies, bringing it into the \nustar\ band. In Fig.~\ref{fig_spin_ratios}, we show the power-law residuals of four simulated spectra with different spin values but otherwise equivalent parameters. These flux changes and energy shifts are clearly visible - the highest spin spectra are both much brighter and extend much further into the \nustar\ band, making it far easier to constrain their spins.

\subsection{The effect of soft X-ray data}
To investigate the effect of including limited soft X-ray data, we simulate a corresponding 5~ks Neil Gehrels \swift\ Observatory X-ray Telescope (\swift\ XRT) snapshot spectrum for each \nustar\ spectrum. The simulated \swift\ XRT spectra are rebinned to a signal-to-noise ratio of 30. We then fit using the same procedure as before. 

\begin{figure*}
\centering
\includegraphics[width=14cm]{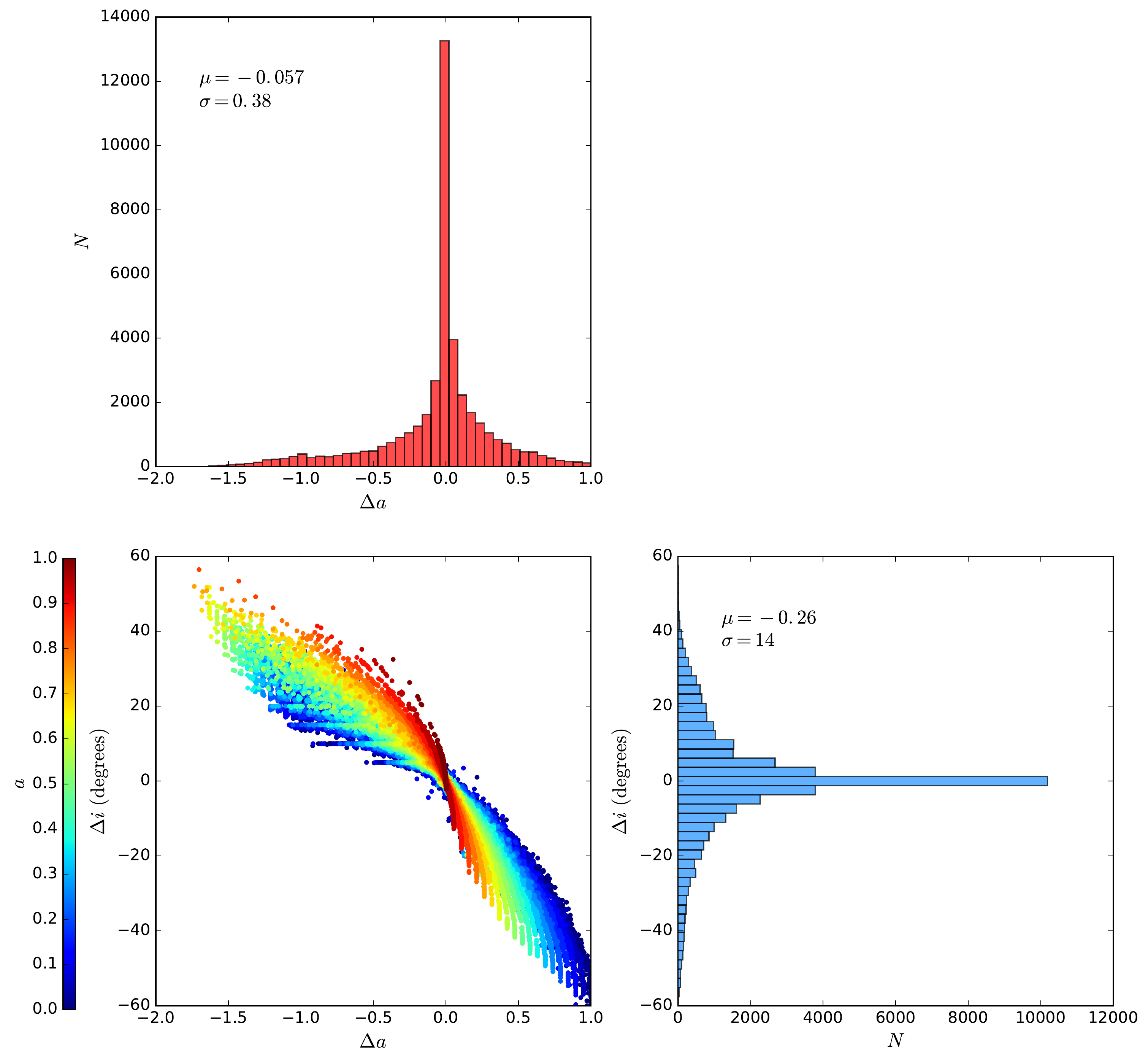}
\caption{The spin-inclination degeneracy for all 41600 spectra in the sample, as in Fig.~\ref{fig_sims_spinincdegeneracy}, but with a 5~ks \swift\ XRT snapshot included along with each \nustar\ exposure. The lower left plot shows the difference between the measured and simulated values of $a$ and $i$, colour-coded by the spin, and the two histograms show the distributions in $\Delta a$ and $\Delta i$ after marginalising over the other parameter. As far fewer fits hit the edge of the parameter space, the peaks visible in the inclination histogram in Fig.~\ref{fig_sims_spinincdegeneracy} are not visible here. Both distributions are more strongly peaked at the correct value ($\Delta a=\Delta i = 0$), and the scatter in the points is much smaller.}
\label{fig_sims_spinincdegeneracy_xrt}
\end{figure*}

The inclusion of XRT data, while it is relatively poor quality compared to the \nustar\ data, makes a significant improvement to the results. In Fig.~\ref{fig_sims_spinincdegeneracy_xrt}, we show the spin/inclination degeneracy for these spectra, which has a much smaller scatter than that in Fig.~\ref{fig_sims_spinincdegeneracy} for the \nustar\ data alone.
Importantly, the inclusion of XRT data also removes the bias in inclination, reducing the mean $\Delta i$ from -6.3\textdegree\ to -0.26\textdegree , well below the measurement error.

We also show the effect of including XRT data on the spin scatter as a function of flux and spin (right panels of Figs.~\ref{fig_flux_dispersion} and \ref{fig_spin_dispersion}, respectively). In both cases, the inclusion of a short XRT exposure greatly improves the results, extending the validity of the method to lower fluxes and lower spin values. 

While these additional spectra contribute very little to the total signal, they allow the peak of the disk spectrum to be accurately measured, which is crucial when the disk spectrum does not extend far enough into the \nustar\ band for this curvature to be seen. This means that the effective temperature can be determined from the \swift\ snapshot, while the flux and the relativistic distortion can be measured by \nustar , giving a significantly more accurate result.

\subsection{Mass and distance}
\label{sec_MD}
Having established that the continuum method can constrain two physical parameters of interest, we now move on to investigating $M$ and $D$ constraints.
As in the case of spin and inclination, there is a strong degeneracy between mass and distance (and accretion rate). In this case, however, there is no extra spectral information to break this degeneracy, and this set of parameters are perfectly degenerate by definition within the model. This means that, contrary to the result of P16, $M$, $D$ and $\dot{M}$ cannot be simultaneously constrained (see \S~\ref{section_discussion} for discussion of this).

\begin{figure}
\centering
\includegraphics[width=\linewidth]{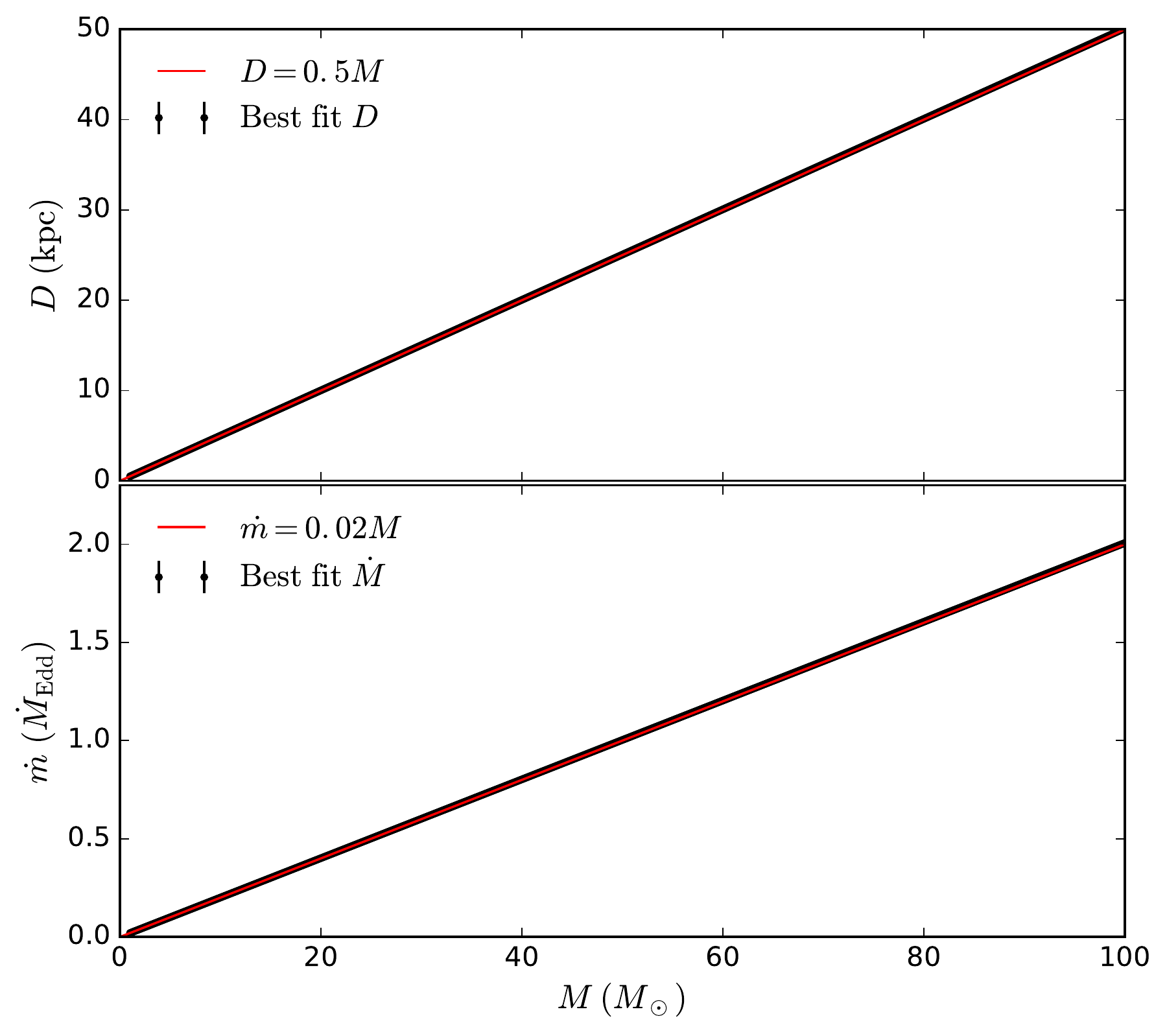}
\caption{Best fit distances and accretion rates as a function of mass for a single \nustar --XRT spectrum. Both are completely degenerate with $M$, but the ratio between the parameters is fixed and very strongly constrained.}
\label{fig_MD_degeneracy}
\end{figure}

To demonstrate this, we consider an individual simulated spectrum ($M=10M_\odot$, $D=5$~kpc, $a=0.9$, $i=60$\textdegree). We fix $a$ and $i$ at the simulated values, then step $M$ from 1--100~$M_\odot$ in 1000 steps, fitting for $D$ at each mass. The results of this are shown in Fig.~\ref{fig_MD_degeneracy}. There is a perfect degeneracy between $M$ and $D$ - the $\chi^2$ value is constant across the whole range of values, so there is no way to distinguish between any two points statistically. This degeneracy is extremely well constrained -the chi-squared increases massively even a small distance from the relation (error bars are shown on the plot, and are much smaller than the points), so the ratio between the parameters is fixed and easily measurable. This means that if any one of the parameters $M$, $D$, or $\dot{M}$ can be constrained by another method, then the other two are immediately constrained to the same degree of accuracy. There are two main scenarios where one of these parameters could be established independently: measuring the distance to the companion, for example using the \emph{Gaia} parallax, or estimating the accretion rate from the hardness-intensity diagram (HID).

\section{Sources of systematic error}
\label{sec_errors}
Our results so far have shown that the $a$ and $i$ can be independently measured, and that the ratio between $M$ and $D$ can be easily and strongly constrained. In both cases, the statistical errors are an accurate reflection of the degree of uncertainty in our simulated data. However, this does not address the likely systematic error, which may be large and even dominate in some circumstances. In this section we attempt to establish some of the largest potential sources of systematic error, where they are particularly applicable to these techniques or have not been extensively discussed by previous authors.

\subsection{Colour-correction factor}
Arguably the largest source of systematic error in continuum fitting measurements is the colour-correction (or spectral hardening) factor, $f$ \citep{Shimura95}. This converts between the colour temperature and the effective temperature of the disk, and depends on the accretion rate and structure of the disk. The colour-correction factor, $f$, is set as 1.7 by default but could be as low as $\sim1.4$ or as high as $\sim2$ \citep{Davis05,Davis06,Reynolds13_ccf}, and is dependent on the mass and accretion rate of the black hole \citep{Davis18}. We test the likely impact of assuming the wrong value of $f$ by fitting the same well-constrained spectrum as before ($M=10M_\odot$, $D=3$~kpc, $a=0.998$, and $i=40$\textdegree), fixing $f=1.4$ and 2.0. This has a moderate impact on the spin and inclination, giving a systematic error of $\sim\pm0.1$ in spin and $\sim\pm10$\textdegree\ in inclination, and a larger effect on the accretion rate, although only in the high $f$ case (the values are given in Table~\ref{table_colourcorrection}). The fit is also much worse in the low $f$ case, but equivalent for high $f$. We also note that the colour-correction factor is meant to account for lower free-free absorption at high energies allowing a deeper view of the disk, but does not necessarily perfectly approximate this effect, which could have implications for precision spectroscopy.

\begin{table}
\centering
\caption{The effect of fitting a well-constrained spectrum with different colour-correction factors. Units are as in Table~\ref{table_wrongspinfits}. The right column gives the standard deviation of the three values.}
\label{table_colourcorrection}
\begin{tabular}{l l l l c}
\hline
Parameter	&	$f=1.4$	&		$f=1.7$	&			$f=2.0$					&	$\sigma$	\\
\hline
$a$			&	$>0.999$		&	$>0.993$	&	$0.81_{-0.06}^{+0.03}$	&	0.1	\\	
$i$			&	$49.4\pm0.1$&	$38.7\pm0.1$&		$51_{-2}^{+4}$			&	6  	\\
$\dot{M}$	&	$0.237\pm0.001$&$0.253\pm0.002$& 	$0.56_{-0.04}^{+0.08}$	&	0.18\\
$\chi^2$	&	585			&	$324$	&			325							\\
\hline
\end{tabular}
\end{table}

\subsection{Calibration Uncertainties}
\label{sec_calibration}
Because the precision needed to make these measurements is extremely high it approaches the limit of the \nustar\ instrumental calibration, which dominates over Poisson noise as the main source of uncertainty in the spectrum at high count rates. This is likely to be an issue for this technique with future instruments as well, although the exact nature will vary, and should in general be carefully considered. The calibration of \nustar\ is discussed extensively in \citet{Madsen15}, so here we focus on two main effects that are particularly relevant to this method: 
\begin{itemize}
\item Uncertainty in column density to the Crab nebula. The nominal column density used in the \nustar\ calibration is $2\times10^{21}$~cm$^{-2}$, but it could be as low as $1\times10^{21}$~cm$^{-2}$ or as high as $6\times10^{21}$~cm$^{-2}$. To estimate the uncertainty introduced by this, we take a NuSTAR spectrum that correctly returns the spin and inclination values ($M=10M_\odot$, $D=3$~kpc, $a=0.998$, $i=40$\textdegree) and fix the column density at values $\pm5\times10^{21}$~cm$^{-2}$ from the simulated value of $5\times10^{21}$~cm$^{-2}$. We find almost no effect on the fit parameters from this, with the largest effect being on inclination, where it adds a systematic error of $\pm3$\textdegree. We conclude that this is not a significant issue, although we note that the effect could be more important in sources where the intrinsic column is low and the uncertainty is thus fractionally larger.
\item Uncertainty in the location of the optical axis. The position of the optical axis is only known to within $\sim30^{\prime\prime}$. The off-axis position is needed for accurate calculation of the effective area, which is crucial for measuring the flux. We test the effect of fitting with an auxilliary response file (ARF) with the wrong off-axis position. We find that, even with the off-axis angle $1^{\prime}$ larger than the true angle, the spin and inclination parameters are not significantly affected. The largest effect is at high energies, where the vignetting effects are strongest \citep[][Fig.~7]{Madsen15} and result in broad line-like features at the $\sim$1--2~\% level in the spectrum. We conclude that this uncertainty in the optical axis is unlikely to have a significant impact on our results.
\end{itemize}
The other main source of uncertainty in the \nustar\ calibration are residuals due to spline fitting to the Crab spectrum. This typically produces narrow line-like features on the order of $\pm$1\%, which should not have a major impact on the results of continuum fitting.

\section{Discussion}
\label{section_discussion}

Arguably the most exciting potential use of being able to measure an additional parameter with continuum fitting is searching for warped accretion disks. A common prediction of accretion simulations is that the disk should bend or tear when the black hole spin axis and the binary axis are misaligned: the material accreted from the companion starts orbiting around the binary axis, but at small radii it aligns with the black hole \citep{Bardeen75}. This is an important tracer of the black hole formation history, as the misalignment is thought to be produced by the supernova `kick' as the black hole is born \citep[see e.g.][and references therein]{Jonker04}.
Because X-ray measurements probe the inner accretion disk, a difference between the inclination measured using X-rays and the binary inclination may indicate a warp in the disk. However, to date all measurements of inner disk inclination with X-ray spectra have been made using only reflection spectroscopy. While there is generally good agreement between the spin values measured with this technique, in some sources there is significant variation in inclination (see e.g. the list of inclinations measured for GX~339-4 listed in the introduction of P16). Inclination measurements with continuum fitting offer an independent method for establishing the inner disk inclination, which shares few systematic uncertainties or modeling assumptions with reflection spectroscopy. If both reflection and continuum methods indicate a different inclination from the binary, we can therefore be reasonably confident in the detection of a warped disk. 

%
\subsection{Additional systematic effects}

%
The main assumption in both continuum and reflection methods is that the disk truncates at the ISCO. If the disk is either truncated or produces significant emission from the plunge region then the estimate of spin from both methods will be wrong. Truncation is not a significant worry in bright states with strong disk emission, and significant truncation can generally be ruled out when the measured spin is high. 
\citet{Zhu12} considered the impact of emission from inside the ISCO on continuum fitting results. They find that the main effect of emission from the plunge region is to modestly raise the measured spin parameter. More worryingly, from our perspective, it also produces a weak power-law tail at high energies. Depending on the photon index of this tail (which is not well established), this could potentially be degenerate with the relativistic distortion of the disk spectrum, compromising the simultaneous spin/inclination measurements. A better understanding of this scattered tail is needed to rule this out.
From the reflection perspective, simulations suggest that emission from the plunge region should be small \citep{Reynolds08}, and contribute less to the total emission the higher the spin is, due to the smaller size of the region. This means that, while still potentially important, the effect of truncation or emission from the plunge region should be small when the measured spin is high. 

We have assumed a simple disk plus power-law model for our spectra. As these measurements rely on high-precision measurements of spectral shape, they are highly sensitive to the exact continuum used. If additional spectral components or spectral complexity \citep[such as the stratified Comptonization discussed in][]{Kawano17} are present in the soft state then these could be highly degenerate with the relativistic distortion of the spectrum. Similarly, overlap with the red wing iron line from relativistic reflection could also produce enough flux on the blue side of the disk spectrum to make such measurements impossible for certain spectra. 

For our simulations we have used a modified version of the \textsc{kerrbb} model, which has been at least partly superseded by \textsc{kerrbb2} and \textsc{bhspec}. We chose to use \textsc{kerrbb} for the sake of simplicity and speed of calculation, as we use a large number of simulated spectra, and because we have only implemented higher-order interpolation in this model (see Appendix~\ref{sec_polykbb}). We also simulate a small number of spectra with the other two models and confirm that the basic principle holds: with high enough signal, both $a$ and $i$ can be constrained.

A flaw in this method as used in P16 was the modelling of the Compton scattered tail. We used the \textsc{comptt} \citep{Titarchuk94} model for the continuum, which does not subtract the photons scattered into the tail from the disk flux, so the photon flux is not conserved. As the continuum-fitting method relies in part upon measuring the total flux, this can potentially lead to an under-estimation of the spin (or, equivalently, an over-estimation of inclination). To avoid this problem, \citet{Steiner09_simpl} created the \emph{simpl} model, which we use for the majority of this work. The model \emph{simpl} self-consistently removes photon flux from the disk spectrum and adds it to a power-law tail, which is cut off at low energies (standard power-law models do not include this, over-predicting the flux at energies below the peak of the disk spectrum). The disadvantage of this model is that it struggles to reproduce more complex continua, which is much more of an issue with \nustar\ than with softer or less sensitive detectors. In P16 we found that \emph{simpl} was not able to reproduce the curvature of the Comptonized component, and relied on \textsc{comptt} instead. Now, however, \citep{Steiner17} have released the \textsc{simplcut} model, which allows the high energy cut off to be modelled either with an exponential function or a physical Comptonization turnover. This model removes the conflict between modelling the high energy spectrum accurately and conserving photon flux.

\subsection{Mass and Distance measurements}
We showed in \S~\ref{sec_MD} that $M$, $D$, and $\dot{m}$ are completely degenerate, but also that their ratios are fixed and strongly constrained. This is contrary to the result of P16, where we claimed a measurement of $M=9.0^{+1.6}_{-1.2}M_\odot$ and $D=8.4\pm0.9$~kpc. The most plausible reason for this disagreement is that we underestimated the error in these parameters in P16, due to not fully exploring the degeneracy between them in our MCMC analysis. This is due to the narrowness of the degeneracy (i.e. the huge increase in $\chi^2$ away from the constant relation), which means MCMC steps will only be accepted if they are almost directly along the line of the degeneracy. This results in a very slow exploration of the parameter space, leading us to conclude that the chains had converged when they had not. However, as the ratios between the three parameters are fixed, we can revise that result to conclude that the true mass and distance should lie close to the line $D/$kpc$=0.93M/M_\odot$. We have since continued the MCMC run for $10^6$ additional steps, and find that it performs as expected: very slowly expanding along the narrow degeneracy. The other parameters are not affected by this, having fully converged in the original run, and are entirely consistent with the values reported in P16.

While we now know that simultaneous constraints on $M$, $D$, and $\dot{m}$ from X-ray spectroscopy alone are not possible, this does not mean that there is no prospect for measuring these values. We need only constrain one parameter through other means to establish the other two to a high degree of precision. For a large number of Galactic XRBs, it is likely that \emph{Gaia} will measure precise distances, without corresponding mass measurements. These distances could then be combined with X-ray spectroscopy to accurately measure the mass. Similarly, with future instruments it will be possible to perform detailed X-ray spectroscopy of extragalactic XRBs, where the distance is known to a high degree of relative precision but the mass cannot be measured dynamically. For these sources, X-ray mass measurements could be a very powerful tool for studying their mass distribution.

Alternatively, we can simultaneously measure $M$ and $D$ if the accretion rate is known. One potential method to establish $\dot{m}$ is to base it on the state transitions transient XRBs go through, which are thought to occur when the accretion rate reaches a certain level \citep[e.g.][]{Maccarone03}. Although the physics here is uncertain, it should be possible to give rough estimates for many transient sources, which can then be used to approximate $M$ and $D$.

A final way of overcoming this difficulty is to use time-resolved spectroscopy. As the mass and difference do not change significantly with time, any two spectra (of the same state) with different luminosities must have different accretion rates. By fitting two or more spectra from the same source simultaneously, it may be possible to constrain the difference in $\dot{m}$, and hence infer $M$ and $D$.

\section{Conclusions}
\label{section_conclusions}
\begin{itemize}
\item \nustar\ spectra can simultaneously constrain both spin and inclination, provided that the source is bright enough and enough of the disk photons are in the \nustar\ band. This conclusion applies in principle to all instruments with sufficient sensitivity and resolution to measure the relativistic distortion of the disk spectrum, which is strongest above 5~keV.
\item These constraints are only strong for bright, high spin sources. For low spins and fluxes, the degeneracy between $a$ and $i$ is unbroken, so the spin is not strongly constrained. This error is well described by the measurement errors, so these measurements should be reliable.
\item The addition of low signal soft X-ray (e.g. \swift\ XRT) data to \nustar\ significantly improves the accuracy of the result. While the XRT does not contribute to measuring the relativistic distortion, it does give an accurate measurement of the peak of the black body flux, which is outside the \nustar\ band in many cases.
\item The mass and distance parameters cannot be simultaneously measured while the accretion rate is free. Fixing $a$ and $i$ to values taken from e.g. reflection spectroscopy and fitting for $M$ and $D$ gives a perfect degeneracy between the two, where the ratio is constant. This degeneracy could be broken by fixing the distance to that measured using \emph{Gaia} or inferring an accretion rate from state transitions. Further into the future, this could be used to measure masses in extragalactic sources, where dynamical masses cannot be measured but the distance is known precisely.
\item We consider the impact of various systematic effects on simultaneous spin and inclination constraints from continuum fitting. We find that the instrumental calibration uncertainties associated with \nustar\ have minimal impact on the results, but that uncertainties in spectral modeling, in particular complex spectral curvature or additional continuum components, can have a major impact on the measured parameters. 
\end{itemize}
Our results show that there is no reason why spin and inclination cannot in principle be measured simultaneously using the disk spectrum in XRBs. There are potentially large systematic errors in the results caused by biases in the spectral fitting, but crucially these are largely independent of the systematic errors in reflection fitting (and also are not unique to this usage of the continuum method). This means that independent measurements of the spin and inclination can be taken with the two methods and used to reliably estimated the inclination of the inner accretion disk, providing a reliable means to search for disk warps.

\section*{Acknowledgements}
MLP and FF are supported by European Space Agency (ESA) Research Fellowships. ACF acknowledges support from ERC Advanced Grant Feedback 340442. DJW acknowledges support form an STFC Ernest Rutherford fellowship.
This work made use of data from the \nustar\ mission, a project led by the California Institute of Technology, managed by the Jet Propulsion Laboratory, and funded by the National Aeronautics and Space Administration. This research has made use of the \nustar\ Data Analysis Software (NuSTARDAS) jointly developed by the ASI Science Data Center (ASDC, Italy) and the California Institute of Technology (USA).
\bibliographystyle{mnras}
\bibliography{bibliography_continuum}

\appendix

\section{polykerrbb}
\label{sec_polykbb}
Both \textsc{kerrbb} and \textsc{bhspec} are calculated on a grid of parameter values, which are then interpolated between. For \textsc{kerrbb}the interpolation is done by the model code, for \textsc{bhspec} the table is interpolated by \textsc{xspec}, in both cases linear interpolation is used. As we have to make high precision measurements of the spectral shape in order to determine both the spin and inclination, the limited resolution of these grids becomes an important factor.

\begin{figure}
\centering
\includegraphics[width=\linewidth]{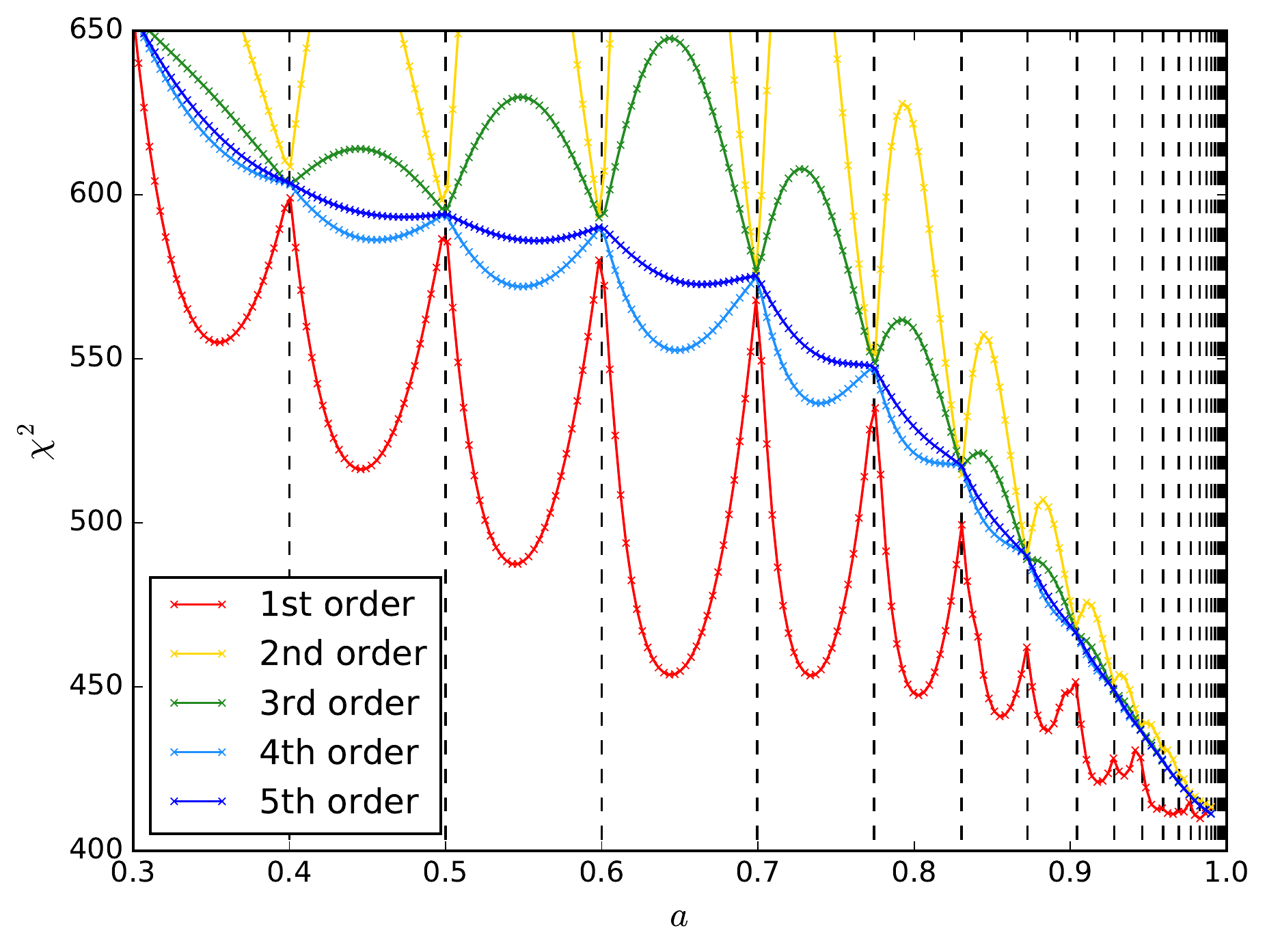}
\caption{$\chi^2$ contour for simulated \textsc{polykerrbb} spectra ($a=0.998$, $i=40$\textdegree, $M=10M_\odot$, $D=3$kpc). Crosses indicate where the fit is evaluated, and vertical dashed lines show the values of spin used in \textsc{kerrbb}. The lowest points (red) are for the 1st order interpolation, and correspond to standard \textsc{kerrbb}.}
\label{fig_kerrbb_contour}
\end{figure}

In Fig.~\ref{fig_kerrbb_contour} we show the $\chi^2$ spin contour for \textsc{polykerrbb} with different degrees of interpolation, for one of the simulated spectra ($a=0.998$, $i=40$\textdegree, $M=10M_\odot$,$D=3$kpc). The grid points where \textsc{kerrbb} is evaluated are marked by the vertical lines, and between the grid points are a series of false minima or maxima, depending on the degree of interpolation. While the best fit is close to the true value in all cases, the fit could easily become trapped in one of these minima as they are both deep and wide compared to the typical step size. The effect of this is twofold: firstly, the fit could get stuck in a false minimum and return the incorrect value, and secondly the errors measured from such a  contour are likely to be significantly under or overestimated\footnote{This has much less effect on results calculated with a fixed inclination, where the constraint on spin is orders of magnitude stronger. Adding the contour with a fixed $i$ value to Fig.~\ref{fig_kerrbb_contour} gives a line that is almost vertical at the correct value.}. Adding a systematic error to the data points scales the whole contour, and does not remove the false minima/maxima.

With a fifth-order polynomial interpolation the false minima are largely removed, although some residuals are still present and do slightly bias our best-fit results towards being found between bins. For our purposes, this model is sufficient, however a more sophisticated spline interpretation would presumably offer an improvement. We note that issues with sparse grid points and linear interpolation in table models are not unique to \textsc{kerrbb} (indeed, \textsc{xspec} exclusively uses linear interpolation for table models), and are likely to be of increasing importance as the sensitivity of X-ray spectra increases. 

The modified \textsc{polykerrbb} code is available from \url{github.com/dbuisson/polykerrbb}. 

\end{document}

%% file: journal_defs.tex
\long\def\symbolfootnote[#1]#2{\begingroup%
\def\thefootnote{\fnsymbol{footnote}}\footnote[#1]{#2}\endgroup} 
\def\araa{ARA\&A}
\def\apj{ApJ}
\def\apjl{ApJ}
\def\apjs{ApJS}
\def\aap{A\&A}
\def\mnras{MNRAS}
\def\pasp{PASP}
\def\pasj{PASJ}
\def\ssr{Space~Sci.~Rev.}